\documentclass[twocolumn]{aastex631}
\pdfoutput=1

\usepackage{amssymb,amsmath}
\usepackage{soul}

\begin{document}

\title{Astrometry as a Tool for Discovering and Weighing Faint Companions to Nearby Stars}

\author[0000-0003-2630-8073]{Timothy D.~Brandt}
\affiliation{Space Telescope Science Institute \\ 3700 San Martin Drive \\ Baltimore, MD 21218, USA}
\affiliation{Department of Physics, University of California, Santa Barbara \\ Broida Hall \\ Santa Barbara, CA, 93106, USA}

\begin{abstract}

This tutorial covers the use of absolute astrometry, in particular from the combination of the Hipparcos and Gaia missions, to identify faint companions to nearby stars and to measure the masses and orbits of those companions.  Absolute astrometry has been used with increasing success to discover new planets and brown dwarfs and to measure masses and orbits for systems with periods as long as centuries.  This tutorial summarizes the nature of the underlying astrometric data, the approach typically used to fit orbits, and the assumptions about that data implicit throughout the process.  It attempts to provide intuition for the sensitivity of astrometry as a function of stellar and companion properties and how the available constraints depend on the character and quantity of data available.  This tutorial is written for someone with some background in astronomy but with no more than a minimal acquaintance with astrometry or orbit fitting.  

\end{abstract}


\section{Introduction} \label{sec:intro}

Astrometry refers to the precise measurement of the positions and motions of the stars and other celestial bodies.  To the ancients, only the known planets could be observed moving across the sky; the stars appeared to be fixed.  As the positions of the planets were measured with increasing precision it drove new models of the Solar system, from the Ptolomeian model to the Copernican model to Kepler's discovery of his laws of planetary motion.  The history of astrometric planet detection dates back to the discovery of Neptune, which was found due to perturbations in the observed orbit of Uranus \citep{LeVerrier_1846}.  

While the motions of planets have long been measured, the motions of stars have a more recent history due the much more difficult nature of that measurement.  Stars undergo annual parallactic motion due to the Earth's changing perspective through the year, but even the closest stars trace parallactic ellipses less than an arcsecond in semimajor axis ($1'' \approx 5 \times 10^{-6}$\,radians), and even the fastest-moving stars travel only a few arcseconds per year.  To measure the parallactic and angular motions of tens of thousands of stars across space, angles across the sky must be measured to milliarcsecond (mas) precision, or $\lesssim$10$^{-8}$\,radians.  This was achieved with the Hipparcos mission \citep{ESA_1997}.  \footnote{For a much more complete history of astrometry, from ancient to modern times, see \cite{Perryman_2012}.  For a very brief history, see \url{https://sci.esa.int/s/8g1qyKw} and associated pages.}

In addition to parallactic and proper motion (motion across the sky), a star with a companion will also orbit the system's center-of-mass.  By measuring this motion we can detect and weigh stellar, substellar, and planetary companions to stars.  Until recently, the required precision was beyond the capabilities of astronomical instruments.  The 2014 launch of the Gaia satellite \citep{Gaia_General_2016}, which has now measured average positions as well as $\sim$20\,$\mu$as ($20\,\mu{\rm as}\approx 10^{-10}$\,radians), makes it possible to measure the reflex motion of stars due to their planetary companions.  

This tutorial covers the use of absolute astrometry---the measurement of positions relative to a fixed reference frame---to discover companions to stars and to fit their orbits and masses.  Section \ref{sec:astrometry_overview} provides an overview of absolute astrometry and astrometric orbital motion.  Section \ref{sec:orbitfit} summarizes the process of fitting an orbit to astrometric data and the attendant assumptions, while Section \ref{sec:datarequirements} summarizes the requirements on an astrometric data set for orbit fitting.  Sections \ref{sec:longperiod} and \ref{sec:shorterperiods} summarize the results and intuition for fitting long- and shorter-period systems, respectively, while Section \ref{sec:companiondiscovery} discusses the use of absolute astrometry to discover new companions to nearby stars.  Section \ref{sec:futureprospects} summarizes future prospects and Section \ref{sec:conclusions} concludes the tutorial.

\section{Overview of Absolute Astrometry} \label{sec:astrometry_overview}

Absolute astrometry consists of position measurements in an inertial reference frame.  All astrometry is, in fact, relative; it only becomes absolute when anchored to objects with accurately known reference positions.  The International Celestial Reference System \citep[ICRS,][]{Charlot+Jacobs+Gordon+etal_2020} represents an ongoing attempt to better realize a set of objects with known absolute positions.  The precise measurements of stellar positions have a wide range of applications in astronomy, from the membership of stellar associations to the distances to stars to orbital motion of stars in multiple systems \citep{Perryman_2012_applications}.

A quasar or other very distant object will remain nearly fixed relative to the ICRS: it is so far away that a physical motion of thousands of km$\,$s$^{-1}$ typically produces angular motion below 1\,$\mu$as\,yr$^{-1}$ ($\lesssim$10$^{-12}$ radians/yr).  In ancient times the stars were viewed as fixed for the same reason: the planets clearly moved across the sky, but the much greater distances to the stars rendered their sky motion unobservable.  With the much higher precisions now possible we can see the motion of the stars.  
Compared to the distant quasars, a star will appear to trace out a sky path defined by position, parallax, and proper motion.  These are referred to as the five basic astrometric parameters \citep{ESA_1997}.  The Hipparcos mission, which observed from 1989-1993 \citep{ESA_1997}, was the first space-based astrometry mission to precisely measure the parallaxes and proper motions of a large sample of stars; it reached precisions of $\sim$1\,mas for about 118,000 stars brighter than $\sim$11th magnitude \citep{ESA_1997}.  The Gaia satellite, in orbit since 2014, can reach precisions almost 100 times better than Hipparcos and measure absolute astrometry for $\sim$1 billion stars brighter than 20th magnitude \citep{Gaia_DR3}.

\begin{figure*}
    \includegraphics[width=\textwidth]{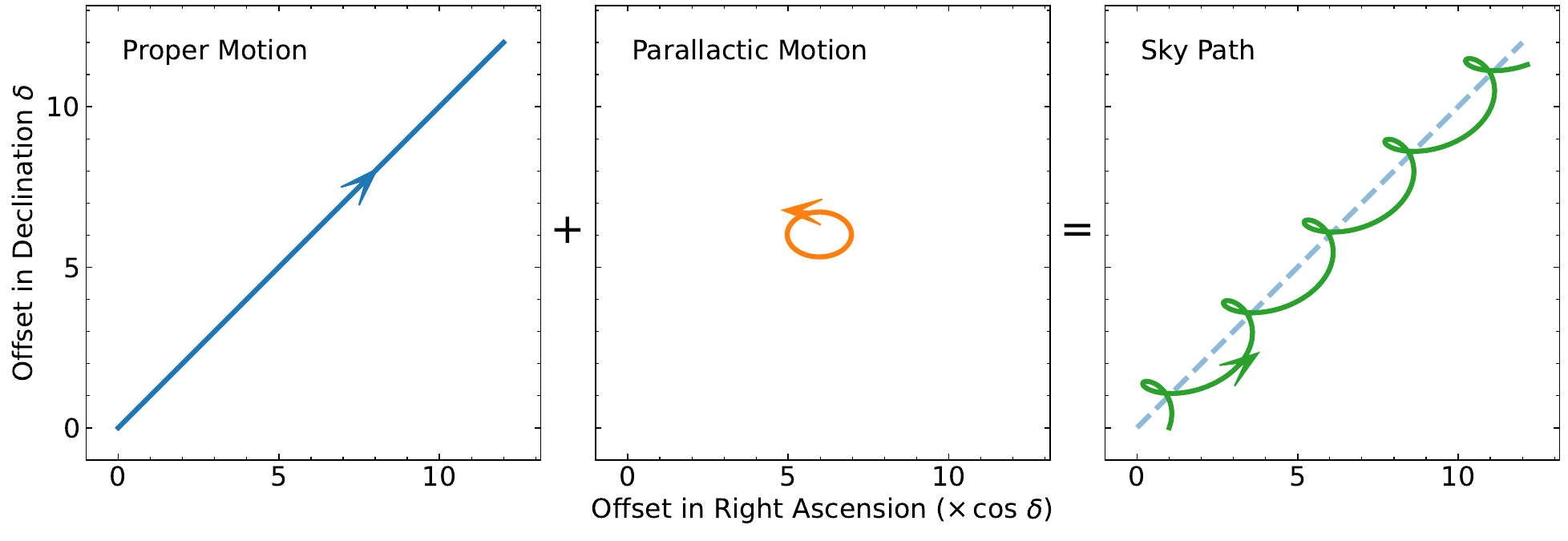}
    \caption{The sky motion of a star moving at constant velocity is the sum of straight-line proper motion (left) and parallactic motion due to the changing perspective of the Earth as it orbits the Sun (middle).  The resulting path, shown on the right, is described by five parameters---two for position, two for proper motion, and one for parallax---and is known as the five-parameter sky path.  Because longitude in spherical coordinates is compressed near the poles, a factor of $\cos \delta$ is commonly multiplied by right ascension so that an arc length has consistent units in the right ascension and declination directions.  Units are arbitrary, but parallactic and proper motion are typically measured in units of mas ($3.6 \times 10^6\,{\rm mas}\,=\,1\,{\rm degree}$) and mas\,yr$^{-1}$, respectively. \label{fig:5parskypath}}
\end{figure*}

Figure \ref{fig:5parskypath} illustrates the sky path followed by a star moving at constant velocity.  The star would appear to move in a straight line if viewed from the barycenter of the Solar system (left panel); this is referred to as proper motion.  Viewed from Earth, the star will appear to move in an ellipse due to the Earth's changing perspective as it orbits the Sun.  This is referred to as parallactic motion, and is shown in the middle panel of Figure \ref{fig:5parskypath}.  The shape of this parallactic motion depends on a star's position in the sky; its amplitude (the size of the ellipse) is referred to as the parallax and depends on the star's distance from Earth.  The right panel of Figure \ref{fig:5parskypath} shows the apparent sky path as the sum of proper and parallactic motion.  

A celestial object moving at constant velocity will, strictly speaking, trace a more complex path in apparent sky coordinates.  Even without parallactic motion, for example, constant linear motion projects nonlinearly onto spherical coordinates.  For most stars, these nonlinear effects are tiny, though many are included in Gaia's data processing \citep{Lindegren+Klioner+Hernandez+etal_2020}.  The deviations from linear apparent motion are small because the changes in angles are typically small: a proper motion of 200\,mas\,yr$^{-1}$, which is larger than that of most stars, is only about $10^{-6}$\,radians\,yr$^{-1}$.  For stars with all but the highest parallaxes and proper motions, the sky paths can be very accurately written as linear functions of position, parallax, and proper motion.  In this limit, a star's positions in right ascension $\alpha$ and declination $\delta$ are given by 
\begin{align}
    \alpha(t) &= \alpha_0 + \mu_\alpha t + \varpi f_\alpha(t) \label{eq:alpha_lin} \\
    \delta(t) &= \delta_0 + \mu_\delta t + \varpi f_\delta(t) \label{eq:delta_lin}
\end{align}
where $\mu_\alpha$ and $\mu_\delta$ are the proper motions in right ascension and declination, and $\varpi$ is the parallax.  The quantities $(\alpha_0, \delta_0)$, termed the reference position, are the coordinates that the star would have at the reference time $t=0$ in the absence of parallactic motion.  The functions $f_\alpha$ and $f_\delta$ may be calculated using the Earth's orbit around the Sun.  They depend on a star's position on the celestial sphere and on the location of an observatory relative to the Solar system barycenter, e.g., near the Earth-Sun-Moon Lagrange point L2 for Gaia \citep{Gaia_General_2016}.  Due to the nature of spherical coordinates, the arclength of a path in right ascension depends both on the difference in angular coordinate and on declination.  An analogous statement holds on the Earth's surface for the same reason: a degree in longitude corresponds to a longer distance near the equator than near a pole.  A $\cos \delta$ term is commonly multiplied by $\alpha$ to produce a quantity sometimes denoted as $\alpha*$, so that an arcsecond in $\alpha*$ has the same arclength as an arcsecond in $\delta$.

Equations \eqref{eq:alpha_lin} and \eqref{eq:delta_lin} describe a path in sky coordinates that depends on the five basic astrometric parameters: $\alpha_0$, $\delta_0$, $\mu_\alpha$, $\mu_\delta$, and $\varpi$.  If the observed positions are not well-fit by a linearized sky path, then this poor fit can be evidence of non-inertial motion, i.e., astrometric acceleration or orbital motion.  Deviations from a linear sky path can also be simply due to the use of spherical coordinates, and can even arise from nonzero radial velocity, as radial motion exchanges with proper motion as distance changes \citep{Lindegren+Dravins_2021}.  These nonlinear effects would not appear if we were to use a Cartesian coordinate system centered on the Solar system barycenter.  Such a coordinate system, however, would be ill-suited to the angular positions that a satellite like Gaia measures.  For most stars, the linear model of Equations \eqref{eq:alpha_lin} and \eqref{eq:delta_lin} is sufficient in the absence of real acceleration, and is appropriate to model precise measurements of angular positions.  Deviations from linear motion then typically reflect orbital motion.  If the orbital period is long then orbital motion appears as a nearly constant acceleration.  For shorter orbital periods, much or all of the star's orbit about its barycenter with a companion may be resolved.  The following sections discuss both of these cases in turn.

Throughout the rest of this tutorial, I will show sky paths without parallactic or proper motion.  That is, I will show the residuals of the actual sky path with respect to the path shown in the right panel of Figure \ref{fig:5parskypath}.  This is for visual clarity: orbital motion is often a tiny perturbation to parallactic and proper motion, and is only detectable statistically or by visually inspecting the residuals from a five-parameter sky path.  Even for very highly significant measurements of astrometric orbital motion, the amplitude of orbital motion can be much less than the system's parallax.

\subsection{Astrometric Acceleration}

Nearly all stars follow a sky path approximately described by the five basic astrometric parameters via Equations \eqref{eq:alpha_lin} and \eqref{eq:delta_lin}.  Superposed on this sky path can be nonlinear or orbital motion, as a star orbits an unseen companion, another luminous star, or even the Milky Way Galaxy.  This section will focus on the limit where this acceleration changes slowly, e.g., for a star in a very long period orbit with a companion.  

The acceleration that a star experiences depends on the mass and separation of its perturber via Newton's law of gravity;  
the acceleration in physical units is given by 
\begin{equation}
    a = \frac{GM}{R^2} . \label{eq:accel_newton}
\end{equation}
However, astrometric missions measure acceleration in angular units rather than physical units, which adds a factor of parallax, and they only measure the acceleration projected onto the plane of the sky.  Placing the acceleration in angular units, 
\begin{equation}
    \frac{\rm acceleration}{0.01\,\rm mas\,yr^{-2}} \approx \left( \frac{M}{M_{\rm Jup}}\right) \times \left( \frac{10\,\rm au}{\rm separation} \right)^2 \times 
    \left(\frac{40\,\rm pc}{\rm distance}\right) .
\end{equation}
The closer a star is to Earth, the more detectable astrometric acceleration becomes.  This differs from the selection effects of other methods of companion detection, like transits or radial velocities.  Both of these are sensitive to properties of the companion like mass, radius, and separation.  Their sensitivity also depends on the star's brightness, as more photons improve the signal-to-noise ratio of flux or radial velocity measurements.  However, the sensitivity of these approaches is otherwise independent of a star's distance from Earth.  

Table \ref{tab:example_accel} shows example astrometric accelerations for a star 10\,pc from Earth under the influence of four perturbers: a giant planet, a low-mass star, a dark matter subhalo, and the Galaxy itself.  For the dark matter subhalo and the Galaxy the mass is only counted if it lies within the star's adopted separation from the barycenter of its perturber.  
The combination of the Hipparcos and Gaia space astrometry missions is now sensitive to astrometric accelerations of $\approx$0.003\,mas\,yr$^{-2}$ for the very best stars \citep{Brandt_2021}, or $\sim$0.01\,mas\,yr$^{-2}$ for a $3\sigma$ detection.  Even this precision may be exceeded when using an ensemble of stars.  The acceleration of the Sun due to the Galaxy has been detected astrometrically as a systematic pattern in quasar proper motions caused by aberration \citep{Gaia_SSAccel}.

Table \ref{tab:example_accel}, together with the current detection limits of $\sim$0.01\,mas\,yr$^{-2}$ for bright stars, shows that perturbations from binary stellar companions can be detected up to a few kpc from Earth, while perturbations from planets can be detected for stars up to $\sim$100\,pc from Earth.  Accelerations from star clusters, molecular clouds, or more exotic sources like dark matter subhalos lie orders of magnitude below current detection limits.

\begin{deluxetable*}{cccc}
    \tablewidth{0pt}
    \tablecaption{Example Astrometric Accelerations for a Star 10 pc from Earth \label{tab:example_accel}}
    \tablehead{Perturber & Mass & Separation & Induced Angular Acceleration}
    \startdata
    Giant planet & 5\,$M_{\rm Jup}$ & 10\,au & 0.2\,mas\,yr$^{-2}$ \\
    Low-mass star & 0.1\,$M_\odot$ & 10\,au & 4\,mas\,yr$^{-2}$ \\
    Dark matter subhalo & $10^4$\,$M_\odot$\tablenotemark{a} & 10\,pc & $10^{-5}$\,mas\,yr$^{-2}$\\
    The Galaxy & $5 \times 10^{10}$\,$M_\odot$\tablenotemark{a} & 8\,kpc & $10^{-4}$\,mas\,yr$^{-2}$ \\
    \enddata
    \tablenotetext{a}{Counting only the mass enclosed by the star's orbit}
\end{deluxetable*}

\subsection{Astrometric Orbital Motion}

Absolute astrometry missions like Gaia measure a star's position at a series of times.  The relevant position is approximately that of the photocenter, the photon-weighted mean position, as long as all photons come from an area that cannot be resolved by the telescope ($\lesssim$$\lambda/D$, or $\lesssim$0$.\!\!''1$ for Gaia with its 1.5\,m primary mirror).  
Orbital motion appears in an astrometric data set when the photocenter is not the same as the system's barycenter.  If two stars have both equal mass and luminosity, then the photocenter and barycenter are identical and no orbital motion will be seen.  In systems of unequal mass, the more massive star will typically be much more luminous.  
As the photons increasingly come from only the more massive star, the photocenter and barycenter become increasingly displaced, and astrometric orbital motion becomes easier to detect.  

In a two-body system, the photocenter will follow a Keplerian orbit.  The semimajor axis of the apparent orbit, assuming all light to come from the primary star, will be the semimajor axis of the mutual orbit multiplied by the ratio of the mass of the faint component to the total mass.  Taking $a$ to be the semimajor axis of the mutual orbit and $M_{\rm sec}$ to refer to the dark component (regardless of whether it has a lower mass), the angular semimajor axis of the photocenter is
\begin{equation}
    a_\phi = \varpi a M_{\rm sec}/M_{\rm tot} .
\end{equation}
The angular size of the apparent orbit could be smaller depending on its orientation: an eccentric orbit aligned with our line-of-sight will have a smaller apparent orbit in the plane of the sky.

The Keplerian motion of the photocenter itself can be described using one of several formulations of the classical orbital elements \citep{Roy_2005}.  A detailed discussion of astrometric orbital modeling and the relevant orbital equations is beyond the scope of this paper, but may be found in, e.g., \cite{Brandt+Dupuy+Li+etal_2021}.  In the following sections I will discuss the overall statistical approach, the assumptions typically made, and the qualitative understanding that a user should have before using orbit fits from Gaia or the outputs of astrometric orbit fitting codes.

\section{Fitting an Orbit} \label{sec:orbitfit}

Measuring a star's astrometric parameters and, if applicable, its acceleration or orbit, requires fitting a sequence of measured positions.  This is always done statistically, typically through $\chi^2$. 
The individual Hipparcos and Gaia measurements are quasi-one-dimensional: their uncertainties are much smaller in one direction than in the orthogonal direction due to each satellite's scanning and observation strategy \citep{ESA_1997,Gaia_General_2016}.  The likelihood of a model astrometric sky path given a sequence of measurements is
\begin{equation}
    -2 \log {\cal L} = \chi^2 = \sum \frac{\left( x_{\rm model} - x_{\rm data} \right)^2}{\sigma^2} \label{eq:chisq}
\end{equation}
where $x$ is the position along the precisely measured direction.  As the satellite continually scans the sky, its orientation changes, and the direction that is precisely measured also changes.  This allows all five of the standard astrometric parameters to be accurately measured.  The astrometric processing pipelines for both Hipparcos and Gaia use Equation \eqref{eq:chisq} to determine their astrometric solutions \citep{ESA_1997, vanLeeuwen_2007,Lindegren+Klioner+Hernandez+etal_2020}.  Because a five-parameter sky path (Equations \eqref{eq:alpha_lin} and \eqref{eq:delta_lin}) is linear in the five basic astrometric parameters, Equation \eqref{eq:chisq}, the log likelihood, is quadratic in these five parameters.  The likelihood itself, $\exp(-\chi^2/2)$ is then a Gaussian whose mean and covariance matrix may be computed using the standard $\chi^2$ machinery.

The form of $\chi^2$ shown in Equation \eqref{eq:chisq} is a special case of the general expression for $\chi^2$; the general form has an inverse covariance matrix in place of the scalar $1/\sigma^2$.  This inverse covariance matrix accounts for correlated errors between the individual astrometric measurements.  An extreme example of a correlated error would be an uncertain overall position, e.g., from an uncertain overall realization of the ICRS.  A resulting error in position would be shared by all astrometric measurements and would not improve as more data are taken. 

A Keplerian orbit can be added to Equations \eqref{eq:alpha_lin} and \eqref{eq:delta_lin}, but the equations are not linear in these additional parameters.  For some formulations of the orbital elements the equations for sky coordinates are linear in a subset of the Keplerian orbital parameters, but never in all of them \citep{Wright+Howard_2009}.  Modeling the sky path is no longer a linear least-squares problem, and it becomes more challenging computationally.  Keplerian orbits are typically fit using a form of Markov Chain Monte Carlo, e.g., in the orbit fitting codes {\tt orbitize!} \citep{Blunt+Wang+Angelo+etal_2020}, {\tt orvara} \citep{Brandt+Dupuy+Li+etal_2021}, and {\tt octofitter} \citep{Thompson+Lawrence+Blakely+etal_2023}.  

The third Gaia data release, DR3, provides Keplerian two-body astrometric orbital solutions for $\sim$10$^5$ stars using only Gaia astrometry \citep{Gaia_DR3}.  These orbital fits adopt the likelihood of Equation \eqref{eq:chisq} combined with a mixture of Monte Carlo and optimization processes \citep{Halbwachs+Pourbaix+Arenou+etal_2023,Holl+Sozzetti+Sahlmann+etal_2023}.  This fit is poorly constrained if the orbital period is long (and hence, if Gaia sees only a small arc of the orbit).  As a result, Gaia orbital searches are limited to periods shorter than twice the DR3 mission baseline, or shorter than about 5.5 years.  An extended Gaia mission will enable the detection of orbits with longer periods \citep{Perryman+Hartman+Bakos+Lindegren_2014}.  

\begin{figure}
    \centering
    \includegraphics[width=\linewidth]{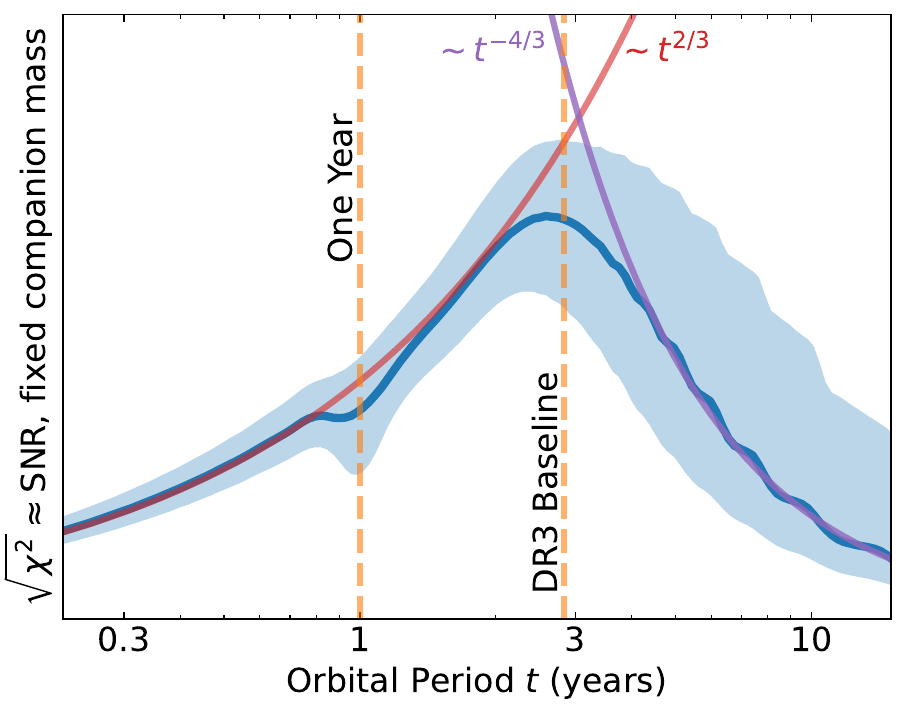}
    \caption{
    Astrometric detectability of a faint companion to a bright star in Gaia DR3 as a function of the system's orbital period, assuming an eccentricity of 0.5.  The shaded region encloses 68\% of orbits with random phases and orientations.  The sensitivity peaks near the Gaia DR3 time baseline, while there is a slight dip around 1 year due to the degeneracy with parallactic motion (a one-year orbit can be indistinguishable from parallactic motion if its eccentricity and orientation are carefully chosen). The vertical scale is arbitrary; it depends on the astrometric precision of Gaia for that star and on the star's distance from Earth. The red and violet lines indicate approximate scalings from Kepler's Third Law and Newton's Law of Gravity for the semimajor axis (red), relevant when a full orbit is measured, and for astrometric acceleration (violet), relevant where only a small orbital arc is measured.  }
    \label{fig:gaia_sensitivity}
\end{figure}

Figure \ref{fig:gaia_sensitivity} shows the astrometric detectability of a faint companion to a visible star in Gaia DR3 as a function of the system's orbital period, assuming an eccentricity of 0.5.  In this case, I take detectable to mean that a fit to the sky path using only parallax, position, and proper motion is insufficient, i.e., the standard $\chi^2$ value of this fit exceeds a certain threshold.  This is the same definition used by \cite{Perryman+Hartman+Bakos+Lindegren_2014}.  The signal-to-noise ratio is similar to the square root of $\chi^2$, which is plotted on the vertical axis.  The minimum detectable mass is proportional to the minimum detectable signal, and inversely proportional to the quantity plotted on the vertical axis.  I further assume that the companion is much less massive than the host star.  With this approximation, a changing companion mass only changes the amplitude of the host star's orbital motion.  

The shaded region of Figure \ref{fig:gaia_sensitivity} encloses 68\% of orbits at random phases and orientations.  The sensitivity initially grows with orbital period.  In this regime the semimajor axis, and thus the astrometric signal, increase with period.  Since I am assuming the companion's mass to be small compared to that of its host, Kepler's Third Law states that the semimajor axis of the star's orbit (and therefore its astrometric detectability) grow as the period to the $\frac{2}{3}$ power; this curve is shown in red. Following a slight dip around a period of 1 year due to degeneracies with parallactic motion, sensitivity increases up to about the Gaia DR3 baseline.  At longer orbital periods the sensitivity decreases because Gaia observes only a fraction of the orbit; it cannot access the increasing astrometric signal.  At longer orbital periods the orbit may not be fully characterized; Gaia may simply measure an astrometric acceleration from a small orbital arc.  Combining Kepler's Third Law with Newton's Law of Gravity yields an acceleration that scales as period to the $-\frac{4}{3}$ power; this curve is shown in violet. As the previous discussion implies, a companion inducing a signal above the detectability threshold of Figure \ref{fig:gaia_sensitivity} can be seen in Gaia DR3 as a deviation from straight-line sky motion, but may or may not permit a good orbital fit.  

While Gaia's sensitivity is limited to modest orbital periods, we can access wider orbits with the aid of additional data.  The Hipparcos mission predates Gaia by about 25 years and, while its measurements are less precise, this long baseline is enough to compensate.  Hipparcos and Gaia combine to measure a long-term proper motion, the difference between the characteristic position during the Hipparcos observations and the characteristic position during the Gaia observations \citep{Brandt_2018,Kervella+Arenou+Mignard+etal_2019}.  This fact formed the basis of the Tycho-Gaia Astrometric Solution, the core of Gaia DR1 \citep{Michalik+Lindegren+Hobbs+etal_2014,Michalik+Lindegren+Hobbs_2015,TGAS_Astrometry_2016}.  The typical uncertainty on this long-term proper motion is the typical uncertainty in a Hipparcos position, anywhere from $\approx$0.5\,mas to 10\,mas, divided by the $\approx$25 years between the observations.  The resulting uncertainties are comparable to uncertainties in Gaia DR3 proper motions \citep{Brandt_2021}.  This enables the difference between the Hipparcos-Gaia mean proper motion and the Gaia proper motion, sometimes referred to as the proper motion anomaly \citep{Kervella+Arenou+Mignard+etal_2019}, to be used as a measure of acceleration.  If other types of measurements, e.g.~radial velocities, are available, then this proper motion anomaly offers a constraint on the orbit.  The underlying assumptions about the astrometric data typically remain that its uncertainties are accurately known and are Gaussian.  A similar $\chi^2$ expression may then be written down and minimized by varying the orbital parameters \citep{Brandt+Dupuy+Bowler_2019}.  

\begin{figure*}
\includegraphics[width=\textwidth]{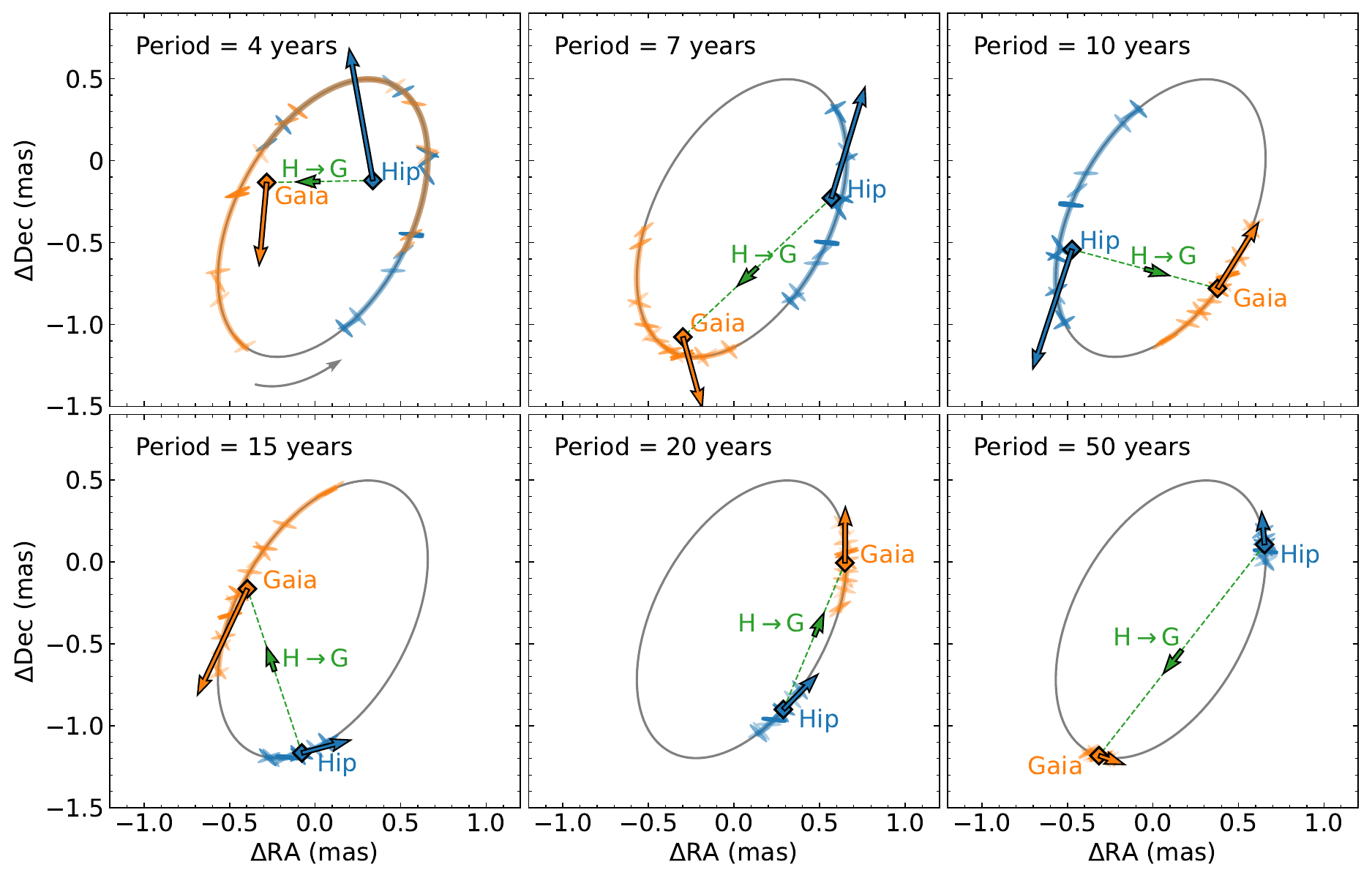}
\caption{Illustration of the positions and proper motions that Hipparcos and Gaia would measure for the same hypothetical orbit but with six different periods.  The orbits shown have had parallactic and proper motion removed (c.f.~Figure \ref{fig:5parskypath}).  The small colored ellipses show the quasi-one-dimensional position measurements made by Hipparcos and Gaia, the diamonds show the best-fit positions at Jyr 1991.25 and 2016.0 from fitting straight lines to the blue and orange points, respectively, and the arrows show the best-fit proper motions.  The portions of the orbit covered by Hipparcos and Gaia are indicated by shading on the orbit, and a black arrow indicates the (counterclockwise) direction of orbital motion.  Differences between the Hipparcos proper motion, the Gaia proper motion, and the scaled difference between the Hipparcos and Gaia positions can constrain orbital motion. At short periods (top middle and especially top left), the positions and proper motions can bear little resemblance to the instantaneous position and proper motion at the average epoch of observation, while at long periods (lower right) this distinction becomes insignificant.  For all periods shown other than 50 years the star completes more than one full orbit between the Hipparcos and Gaia central epochs. \label{fig:propermotion_demo}}
\end{figure*}

Figure \ref{fig:propermotion_demo} shows the actual sky path taken by a star for a sample orbit at six different periods.  The sky paths shown have had parallactic and proper motion removed to make the orbital motion easier to see.  The angular semimajor axis is the same in all of these examples.  Maintaining the same orbital period would require a higher system mass at shorter orbital periods.  

Short orbital periods are accessible to Hipparcos or Gaia individually; each mission covers most or all of an orbit.  These short orbital periods are more difficult to constrain with only the catalog proper motions and a long-term proper motion.  The positions and proper motions obtained from fitting Equations \eqref{eq:alpha_lin} and \eqref{eq:delta_lin} bear little visual correspondence with the orbit at these short orbital periods (top-middle, and especially top-left, panels of Figure \ref{fig:propermotion_demo}).  As the orbital period becomes comparable to or longer than the time baseline between Hipparcos and Gaia, the three proper motions measure something close to an instantaneous acceleration.  In this limit, most nearly shown in the lower-right panel of Figure \ref{fig:propermotion_demo}, the combination of Hipparcos and Gaia holds much more power to constrain an orbit than Gaia alone.

\subsection{Hipparcos-Gaia Accelerations vs.~Gaia Two-Body Orbits}

This tutorial primarily focuses on using Hipparcos and Gaia positions and proper motions.  With Gaia DR3 \citep{Gaia_DR3}, two-body orbital fits are also available for a number of stars \citep{Halbwachs+Pourbaix+Arenou+etal_2023,Holl+Sozzetti+Sahlmann+etal_2023}.  These two-body fits include a set of seven Keplerian orbital parameters sufficient to fully describe the orbit of the photocenter.  Gaia DR3 includes a covariance matrix for these seven parameters.  Their joint distribution will only be Gaussian, however, in the high signal-to-noise ratio limit.  For example, orbital phase is periodic, eccentricity and period are positive, and eccentricity is bounded; the likelihood cannot be Gaussian in these parameters unless they are very tightly constrained.  This is in contrast to the five basic astrometric parameters\footnote{While physically meaningful parallaxes are positive, a negative parallax would describe a valid sky path.  Negative parallax values are not mathematically excluded when fitting sky paths and do appear in Hipparcos and Gaia data releases.}.  The fact that a sky path is linear in the five basic parameters of position, proper motion, and parallax means that their posterior distribution will indeed be Gaussian assuming the data to have Gaussian errors.  

Additional data in the form of direct imaging or radial velocities can improve a Gaia two-body orbit and can measure the difference between photocenter motion and the separation between the primary star and its companion.  When using the Gaia two-body orbit together with additional data, the limitation of Gaia's Gaussian approximation to the joint likelihood of the orbital parameters can become important.  In this case, it is important to test the consistency of the RVs and/or imaging data with the Gaia solution.  This is possible in a few cases.  For example, for Gaia BH-1, the RV orbit is inconsistent with the Gaia two-body fit unless the Gaia uncertainties are inflated by a factor of $\approx$2 \citep{ElBadry+Rix+Quataert+etal_2023,Chakrabarti+Simon+Craig+etal_2023}.  

The individual Gaia astrometric measurements have not been released as of the time of writing, making it impossible to perform orbital fitting to Gaia data independently of the Gaia Data Processing Consortium.  The individual measurements will be released in Gaia DR4.  Gaia DR3's formal uncertainties on the astrometric parameters are known to underestimate the true uncertainties by up to $\approx$30\%-40\% for bright stars \citep{El-Badry+Rix+Heintz_2021,Brandt_2021}.  This could impact the treatment of measurements and uncertainties in orbit fitting and, by extension, the values, uncertainties, and biases of the resulting orbital parameters.  
With future Gaia data releases, fits to the Gaia two-body solutions will not be necessary; fits will be possible on the individual astrometric measurements.  As noted in the following section, using these individual measurements in practice will require a detailed understanding of their statistical properties.

\subsection{Example: Fitting Individual Gaia Measurements}

While the individual Gaia position measurements will not be available until DR4, the full time series was recently released for a visible star hosting a dark, almost certainly black hole, companion \citep{Gaia_BH3_2024}.  The model those authors fit to the individual astrometric measurements is that of a single star (Equations \eqref{eq:alpha_lin} and \eqref{eq:delta_lin}) plus a Keplerian orbit in the plane of the sky \citep[e.g.][]{Roy_2005}.  The astrometric measurements are precise only in one direction (the scan direction), so the model astrometric path is the sum of the projections of right ascension and declination onto this scan direction.  

With the full orbital model plus single star sky path, the right ascension and declination will be, using the Thiele-Innes representation (see, e.g., \cite{Wright+Howard_2009} for definitions of quantities),
\begin{align}
    \alpha &= \alpha_0 + \mu_\alpha t + \varpi f_\alpha + A X(e, T_0, P, t) + F Y(e, T_0, P, t) \\
    \delta &= \delta_0 + \mu_\delta t + \varpi f_\delta + B X(e, T_0, P, t) + G Y(e, T_0, P, t)
\end{align}
or, in the Campbell representation,
\begin{align}
    \alpha &= \alpha_0 + \mu_\alpha t + \varpi f_\alpha + R(e, T_0, P, i, \omega, \Omega, a, t) \\
    \delta &= \delta_0 + \mu_\delta t + \varpi f_\delta + S(e, T_0, P, i, \omega, \Omega, a, t).
\end{align}
Assuming a scan angle $\phi_i$ and a measured position of $y_i$ along the scan direction for measurement $i$, the $\chi^2$ value representing the log of the likelihood is
\begin{equation}
    \chi^2 = \sum_i \frac{\left( \alpha_i \sin \phi + \delta_i \cos \phi - y_i \right)^2}{\sigma_i^2}
\end{equation}
where $\sigma_i$ is the uncertainty of measurement $i$.  Gaia-BH3 also has radial velocity data; the total log likelihood is then the sum of the $\chi^2$ given above along with 
\begin{equation}
    \chi^2_{\rm RV} = \sum_i \frac{\left( {\rm RV}_{{\rm model},i} - {\rm RV}_i \right)^2}{\sigma_i^2},
\end{equation}
where the model radial velocities ${\rm RV}_{{\rm model},i}$ depend on the orbital elements \citep[e.g.][]{Wright+Howard_2009}, ${\rm RV}_i$ are the measured radial velocities, and $\sigma_i$ are their uncertainties.  The total $\chi^2$ may then be minimized using standard techniques, or treated as a likelihood using ${\cal L} \propto \exp\left[-\chi^2/2\right]$ and treated using tools like Markov Chain Monte Carlo.  An example Jupyter notebook provided by the Gaia-BH3 discovery team illustrates the approach and the resulting fit\footnote{\url{https://github.com/esa/gaia-bhthree}}.  

There are many subtleties to fitting such an orbit that are well beyond the scope of the present discussion; I list a few of them here.  The formulations with Thiele-Innes and Campbell elements may not be equivalent, as the Jacobean determinant of the transformation from one to the other is not unity.  The approach outlined here also assumes independent, Gaussian errors for the Gaia measurements; this must be rigorously checked.  Outliers in the data must be rejected, but this can be handled in different ways and, in any case, assumes the validity of the Gaussian model for the bulk of the data.  Finally, the likelihood will not be Gaussian, leading to subtleties in the interpretation of orbital constraints for large samples.  

\section{Requirements on an Astrometry Data Set} \label{sec:datarequirements}

The preceding section presented an overview of the process of using absolute astrometry to fit orbits.  Whether using a sequence of astrometric measurements as Gaia DR3 does for its orbital solutions \citep{Halbwachs+Pourbaix+Arenou+etal_2023,Holl+Sozzetti+Sahlmann+etal_2023}, or whether using just two or three proper motion measurements from combining Hipparcos and Gaia data \citep{Brandt+Dupuy+Bowler_2019}, an orbital fit presumes the validity of Equation \eqref{eq:chisq}.  

The use of $\chi^2$ statistics assumes that all measurements are drawn from Gaussian distributions where the mean is the true value and the standard deviation is typically known (and identified with the measurement uncertainty).  In order for this assumption to be satisfied the data must be free of systematics and and have accurately calibrated uncertainties.  This latter requirement extends to the covariance.  If the data have unmodeled covariance, the resulting parameter uncertainties may be severely underestimated even while the astrometric measurements are scattered about the best-fit sky path as expected.  As an example, consider a series of perfectly covariant measurements.  This means that a measurement error in one measurement will be repeated identically in all others; the measurements will show no scatter relative to one another.  If we estimate the uncertainty from the difference between the various measurements, we will miss their shared error.  In astrometry, a shared but unmodeled error could appear to be an unexpected offset in position on the sky.  The residuals of the measurements from the (incorrect) best-fit sky positions could still be near zero.  An unmodeled error in position, however, would have a decisive impact on the measurement of long-term proper motions measured between astrometric missions.

The issues outlined above apply to the five-parameter astrometric fits (parallax, position, proper motion) of Hipparcos and Gaia just as well as they apply to orbital fits.  In order to use these quantities, we must verify their consistency and statistical properties.  Fortunately, consistency checks are available on the statistical properties of the five-parameter fits.  In the case of Hipparcos, these may be done, e.g., with negative derived parallaxes.  Negative parallaxes are unphysical.  However, a linear least-squares fit to a very distant object will return a negative parallax half the time.  The distribution of negative parallaxes was used to validate the new Hipparcos reduction \citep{vanLeeuwen_2007}.  For Gaia, the same test is available for the parallaxes of objects known to be quasars \citep{Babusiaux+Fabricius+Khanna+etal_2023}.  These checks are useful but do not apply to all data.  For example, while there are hundreds of thousands of quasars in Gaia, none of them are as bright as the stars of Hipparcos.  
For these bright stars in Gaia, we may perform consistency checks using wide binaries (that should have nearly the same parallax and proper motion) \citep{El-Badry+Rix+Heintz_2021,Cantat-Gaudin+Brandt_2021}.  

Measurement uncertainties may be estimated using knowledge of the instrument and spacecraft as well as the scatter of measurements about the best-fit sky paths.  If there are any unmodeled covariances between measurements, this process may result in underestimated uncertainties in the resulting astrometric parameters.  This represents the difference between what may be termed the internal, or formal, uncertainties, and the total, or internal+external, uncertainties \citep{Arenou+Luri+Babusiaux+etal_2017,Arenou+Luri+Babusiaux+etal_2018,Fabricius+Luri+Arenou+etal_2020}.  The Hipparcos re-reduction incorporates a small, additional error term added in quadrature with all of the individual astrometric measurements \citep{vanLeeuwen_2007}.  The original Hipparcos catalog \citep{ESA_1997} did not specifically tally the external error, but estimated the total error to be a factor of up to $\approx$20\% larger than the catalog uncertainty \citep{Perryman+Lindegren+Kovalevsky+etal_1997}.  \cite{Brandt+Michalik+Brandt_2023} find that the Hipparcos-2 uncertainties likely remain underestimated for bright stars, with uncertainties underestimated by as much as a factor of $\approx$4.  For Gaia, the ratio of total to internal uncertainty varies as a function of magnitude.  At relatively faint magnitudes, parallaxes and proper motions of quasars and the consistency of astrometric fits for so-called duplicated sources find that Gaia uncertainties are underestimated by $\approx$5-10\% depending on the catalog version \citep{Arenou+Luri+Babusiaux+etal_2017,Arenou+Luri+Babusiaux+etal_2018,Fabricius+Luri+Arenou+etal_2020}.  For bright stars the formal Gaia DR3 uncertainties underestimate the true uncertainties by factor of $\approx$1.3-1.4 as measured from wide binaries \citep{El-Badry+Rix+Heintz_2021} and from the consistency of Hipparcos and Gaia \citep{Brandt_2021}.  Underestimated uncertainties and systematics can also be a function of magnitude: Gaia DR3 has residual systematics in the proper motions of stars just brighter than magnitude 13, where the data collection method (the gating) changes to avoid saturation \citep{Cantat-Gaudin+Brandt_2021}.  

The need to inflate uncertainties above their internal values raises questions about the suitability of these data for orbit fitting.  For the five astrometric parameters, we may at present only calibrate the final values and uncertainties of the fitted parameters, not the individual position measurements that were used in the fit.  In some cases, similar tests with external data are possible for orbital motion, e.g., where long-term radial velocity monitoring is available.  There are a handful of examples in the literature, among them Gaia-BH-1 \citep{ElBadry+Rix+Quataert+etal_2023,Chakrabarti+Simon+Craig+etal_2023}.  In this case, RV monitoring of the visible primary star yields an orbit marginally inconsistent with the Gaia two-body fit.  The error inflation needed to achieve agreement with the RV parameters is a factor of $\approx$2, substantially larger than the factor needed to account for external errors for stars of similar brightness \citep{El-Badry+Rix+Heintz_2021}.  

The results outlined above suggest that orbit fitting should be done with caution unless a data set has been specifically calibrated for that purpose.  Work to understand the individual Hipparcos position measurements at this level is ongoing.  For many Hipparcos stars, Gaia has two-body orbital fits, and we can check the consistency of the Hipparcos measurements with these Gaia sky paths.  Existing work by \cite{Brandt+Michalik+Brandt_2023} suggests that the Hipparcos re-reduction should be treated with caution due to its use of the bright stars' residuals to five-parameter sky paths in order to calibrate the spacecraft attitude.

Though individual Gaia astrometric measurements are not yet available and questions linger about the use of Hipparcos intermediate data, it is possible to cross-calibrate and use the Hipparcos and Gaia {\it catalogs} for orbit fitting.  In order to use Hipparcos and Gaia data to fit orbits, both must represent positions and proper motions in the same inertial reference frame.  Gaia for bright stars is, in fact, calibrated to Hipparcos via the long-term proper motion \citep{Lindegren+Klioner+Hernandez+etal_2020}.  The consistency of Hipparcos and Gaia can be tested assuming that most stars have no detectable astrometric acceleration.  For these stars, the $z$-score, the difference between the Gaia and the long-term proper motion normalized by their combined uncertainty, should be distributed as a unit Gaussian:
\begin{equation}
    z = \frac{\mu_{\rm HG} - \mu_{\rm G}}{\sqrt{\sigma^2_{\rm HG} + \sigma^2_{\rm G}}}
\end{equation}
This test was used to construct and validate the Hipparcos-Gaia Catalog of Accelerations \citep{Brandt_2018,Brandt_2021}.  The following sections summarize the combined use of Hipparcos and Gaia to identify massive companions to nearby stars, to fit their orbits, and to measure their masses.

\section{Masses and Orbital Constraints: Long Period Limit} \label{sec:longperiod}

In the limit of an orbital period much longer than the time baseline between Hipparcos and Gaia, absolute astrometry only probes a small fraction of the orbital arc.  In this limit Hipparcos and Gaia measure an acceleration in the plane of the sky.  According to Newton's law of gravity, the large physical separation required to have a $\sim$centuries-long period results in a low acceleration.  In order for the astrometric acceleration to be detectable, circumstances must be favorable.  These systems are easier to detect when they are nearby (so that the same astrometric acceleration in physical units is larger when projected onto the sky in angular units).  They are also easier to measure for more massive companions, which exert larger tugs on their host stars.  Nearby binary systems with a stellar, stellar remnant, or brown dwarf companion are ideally suited to the measurement of astrometric accelerations.  Substellar examples include Gl~229B (5.8\,pc, 71\,$M_{\rm Jup}$) and Gl~758B (15.6 pc, 38\,$M_{\rm Jup}$) \citep{BrandtGM+Dupuy+Li+etal_2021}, white dwarf examples include Gl~86B \citep[10.8\,pc, 0.61\,$M_\odot$,][]{Zeng+Brandt+Li+etal_2022}, while there is a large and increasing number of stellar companions that permit these dynamical mass measurements \citep[e.g.][]{Rickman+Matthews+Ceva+etal_2022}.  

The examples cited above have bright primary stars and faint companions.  In this limit Hipparcos and Gaia measure the location of the star, because that is the source of nearly all of the light.  If the companion emits a significant fraction of the system's light, then Hipparcos and Gaia measure the motion of something closer to the system's photocenter and the analysis becomes more complicated.  In this section I will assume that all photons come from the primary star.   The star's astrometric acceleration, which I denote $a_\perp$ (because it is perpendicular to the line of sight), is related to the companion mass $m_2$ by
\begin{equation}
    a_\perp = \frac{Gm_2}{r^2} \sin \phi \label{eq:a_perp}
\end{equation}
where $r$ is the physical separation between the two bodies and $\phi$ is the angle of the two bodies' separation vector with respect to the line of sight.  While $a_\perp$ may be directly measured by Hipparcos and Gaia, a constraint on the companion mass independently of separation is not possible without additional information.  

For many nearby binary systems, however, additional information is available in the form of a radial velocity trend and the separation of the two bodies in the plane of the sky.  The radial velocity trend measures the line-of-sight acceleration $a_\parallel$,
\begin{equation}
    a_\parallel = \frac{Gm_2}{r^2} \cos \phi . \label{eq:a_parallel}
\end{equation}
This combines with $a_\perp$ to measure the three-dimensional acceleration in an inertial reference frame (the Solar system barycenter provides a very nearly inertial reference frame for radial velocities).  Equations \eqref{eq:a_perp} and \eqref{eq:a_parallel} can directly constrain $m_2/r^2$, but cannot constrain either $m_2$ or $r$ individually.  To break this degeneracy we need one more measurement, which can be supplied by the two bodies' projected separation.  This may be measured from high-contrast imaging, speckle imaging, or, for very nearby binary systems, the two bodies may be separately resolved within Gaia itself.  The projected separation $\rho$ is given by 
\begin{equation}
    \rho = r \sin \phi . \label{eq:proj_sep}
\end{equation}
In the limit of a long orbital period and nearly simultaneous measurements of $a_\parallel$, $a_\perp$, and $\rho$, Equations \eqref{eq:a_perp}, \eqref{eq:a_parallel}, and \eqref{eq:proj_sep} may be combined to solve for the companion mass $m_2$ \citep{Brandt+Dupuy+Bowler_2019}.  

\begin{deluxetable*}{ccccccc}
    \tablewidth{0pt}
    \tablecaption{Nearby Long-Period Systems with Precise Dynamical Masses \label{tab:nearby_dynamical_masses}}
    \tablehead{System & Companion Type & Distance (pc) & Period (yr) & $\Delta\mu$ Significance\tablenotemark{a} & Companion Mass & Reference}
    \startdata
    Gl~86 & White Dwarf & 10.8 & $97 \pm 2$ & $281\sigma$ & $0.543 \pm 0.004$\,$M_\odot$ & \cite{Zeng+Brandt+Li+etal_2022} \\
    HD~114174 & White Dwarf & 26.4 & $104 \pm 5$ & $88\sigma$ & $0.59 \pm 0.01$\,$M_\odot$ & \cite{Zhang+Brandt+Kiman+etal_2023} \\
    HD~159062 & White Dwarf & 21.6 & $387^{+60}_{-73}$ & $99\sigma$ & $0.61 \pm 0.01$\,$M_\odot$ & \cite{Bowler+Cochran+Endl+etal_2020} \\
    HIP~22059 & Low-mass Star & 30.9 & $93 \pm 6$ & $208\sigma$ & $0.237 \pm 0.004$\,$M_\odot$ & \cite{Rickman+Matthews+Ceva+etal_2022} \\
    HD~157338 & Low-mass Star & 33.0 & $126^{+34}_{-21}$ & $135
    \sigma$ & $0.34 \pm 0.01$\,$M_\odot$ & \cite{Rickman+Matthews+Ceva+etal_2022} \\
    Gl~229 & Brown Dwarf & 5.8 & $238 \pm 5$ & $114\sigma$ & $71.4 \pm 0.6$\,$M_{\rm Jup}$ & \cite{BrandtGM+Dupuy+Li+etal_2021} \\
    Gl~758 & Brown Dwarf & 15.6 & $154^{+63}_{-39}$ & $41\sigma$ & $38.0 \pm 0.7$\,$M_{\rm Jup}$ & \cite{BrandtGM+Dupuy+Li+etal_2021} \\
    \enddata
    \tablenotetext{a}{Approximate significance (in Gaussian sigma) of non-constant proper motion between Hipparcos and Gaia}
\end{deluxetable*}

Table \ref{tab:nearby_dynamical_masses} lists several nearby systems with orbital periods $\gtrsim$100 years, with RV trends, relative astrometry, and astrometric accelerations.  These systems provide some of the best first-principles mass measurements of substellar objects, stellar remnants, and very low-mass stars and help anchor models of their evolution.  The precise mass measurements listed in Table \ref{tab:nearby_dynamical_masses} would not be possible without the highly significant measured astrometric accelerations.  With only radial velocity measurements or a relative orbit on the sky, the total mass is typically inferred from the period and semimajor axis via Kepler's Third Law, which in turn requires a good measurement of the period.  This typically means that the data must cover most or all of an orbital period.  Very few systems with orbital periods of a century or more have meaningful data extending back that long.  Absolute astrometry, by measuring an acceleration in an inertial reference frame, entirely removes the requirement to cover most of an orbital period in order to apply Kepler's Third Law.  Even with a short orbital arc, it is possible to apply Newton's Second Law instead to infer a companion mass directly.

\section{Masses and Orbital Constraints: Shorter Periods} \label{sec:shorterperiods}

At shorter orbital periods, some systems have a sufficiently large and well-resolved astrometric signal that Gaia DR3 provides two-body orbital fits.  Many more do not.  These systems, like the long-period systems discussed in the previous section, have three proper motions between the Hipparcos and Gaia catalogs.  The differences between these three proper motions do not measure something close to the instantaneous acceleration of the star, but something more complex, as indicated in the top row of panels of Figure \ref{fig:propermotion_demo}.  

Just as in the long-period limit, the three proper motions from Hipparcos and Gaia only offer a joint constraint on mass and separation.  An important difference from the long period limit is that radial velocity and relative astrometric monitoring by direct imaging are much more likely to measure orbital curvature.  This both reduces the need to have both of these types of measurements and increases the importance of detailed orbital modeling.  For example, excellent constraints on companion mass can be achieved with absolute astrometry and radial velocity, so long as the radial velocity measurements cover a large fraction of an orbit.  Examples of these measurements may be found in \cite{Xuan+Wyatt_2020}, \cite{Li+Brandt+Brandt+etal_2021}, \cite{Hill+Kane+Campante+etal_2021}, and \cite{Venner+Vanderburg+Pearce_2021}.  If direct imaging is also available, the constraints can be extremely tight \citep[e.g.][]{Balmer+Pueyo+Stolker+etal_2023}.

As the orbital period decreases, the measured proper motions become increasingly sensitive to precisely when and in which direction Hipparcos and Gaia measured the individual positions.  For Hipparcos these times and measurement directions are available via the intermediate astrometric data.  For Gaia they are not, but the Gaia Observation Scheduling Tool \citep[GOST,][]{gaia_gost_user_manual} includes predictions of the epochs and scan directions.  Some measurements will inevitably be more precise than others, and some observations may be unusable.  As a result, GOST is an imperfect substitute for the observations themselves.  

The most correct way to treat the data at shorter orbital periods is to calibrate the uncertainties and covariances of the individual position measurements and then to use them directly.  This is not possible now, and could be difficult well into the future.  As the orbital period approaches the time baseline of either Hipparcos or Gaia, i.e.~$\lesssim$10\,years, orbital fits using only the catalog proper motions become increasingly sensitive to the details of the assumed astrometric measurements.  These can be checked to a limited degree.  The Hipparcos-Gaia Catalog of Accelerations provides the characteristic observational epochs of Hipparcos and Gaia in right ascension and declination, and these can be checked against the mean epochs of the assumed measurements, e.g.~from GOST, in each direction.  Whether or not this checks out, orbital fits using the Hipparcos-Gaia Catalog of Accelerations or a similar astrometric catalog should be evaluated with caution at periods $\lesssim$10\,years, and especially so at periods $\lesssim$5\,years.  

\subsection{Many-body systems}

If a star has more than one companion, its astrometric and radial velocity motions will be the sum of the contributions of each companion.  This is now treated in orbit-fitting codes using a superposition of Keplerian orbits \citep[e.g.][]{Blunt+Wang+Angelo+etal_2020,Brandt+Dupuy+Li+etal_2021}.  Some care must be taken with the relative astrometry.  Each body will follow an approximately Keplerian orbit around the barycenter of all bodies interior to its orbit (including itself).  This means that the relative orbit of an outer planet around its host star will differ from an ellipse \citep{Lacour+Wang+Rodet+etal_2021}.  As precisions improve, especially due to the GRAVITY instrument on the Very Large Telescope \citep{GRAVITY_2017}, a superposition of Keplerian orbits will become insufficient to describe the dynamics and full $N$-body modeling will become necessary \citep{Covarrubias+Blunt+Wang_2022}.  For now, however, Keplerian orbits are sufficient to describe the dynamics of systems like $\beta$~Pictoris and HD 206893 to derive the masses of their planetary companions \citep{Brandt+Brandt+Dupuy+etal_2021,Hinkley+Lacour+Marleau+etal_2023}.  

\section{Companion discovery from astrometric acceleration} \label{sec:companiondiscovery}

Astrometric accelerations are measured by Hipparcos and Gaia for nearly all bright stars across the sky.  This raises the possibility of using absolute astrometry to identify previously unknown dark companions.  An astrometrically accelerating star is being tugged by an unseen companion; the magnitude of the acceleration constrains the companion's separation and mass.  This fact may be used to select targets for follow-up observations, either in radial velocities or in direct imaging, to confirm these companions, to weigh them, and to characterize their atmospheres.  

In the remainder of this section I will assume that the companion is faint, i.e., Hipparcos and Gaia measure only the motion of the primary star.  If the companion is comparably luminous to its host star, Hipparcos and Gaia may instead trace the path of their photocenter and underestimate the motion of the primary star itself.  In the limit that the primary and secondary stars have equal brightness no astrometric signal will be measured at all.

\begin{figure*}
    \includegraphics[width=0.5\textwidth]{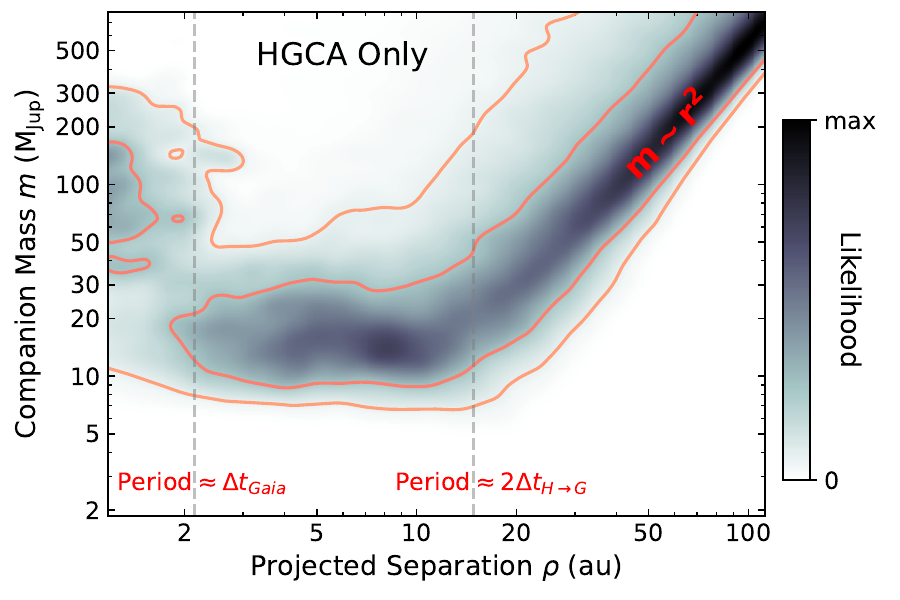}
    \includegraphics[width=0.5\textwidth]{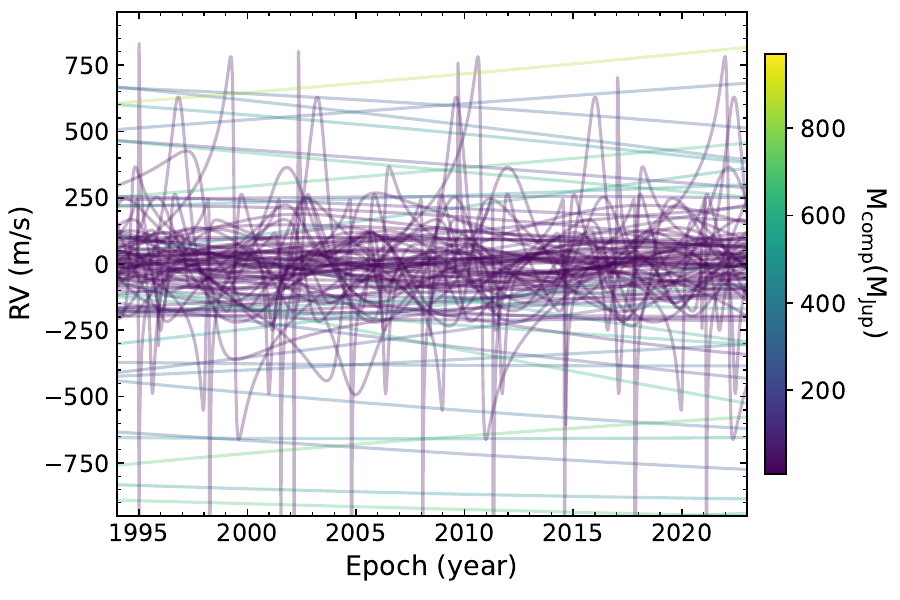}
    \includegraphics[width=0.5\textwidth]{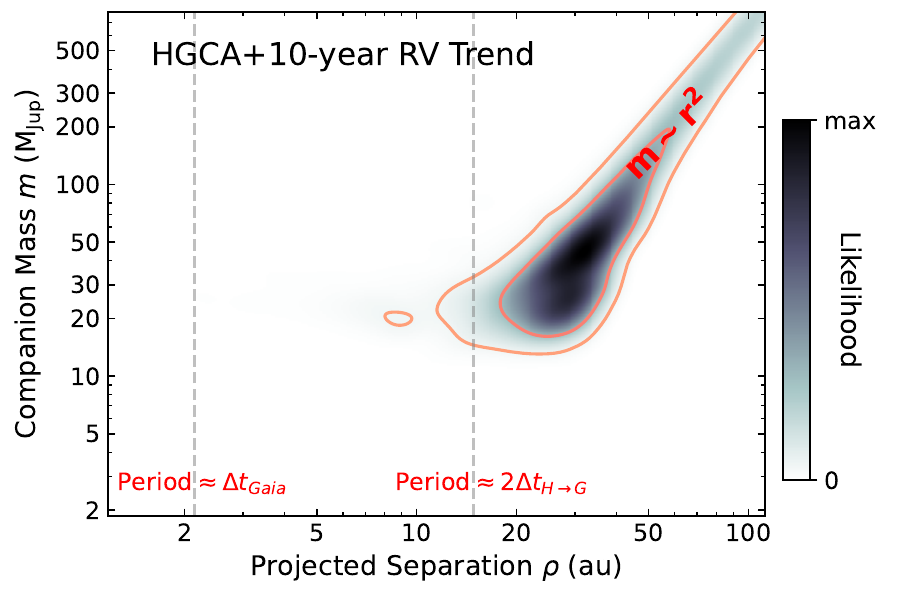}
    \includegraphics[width=0.5\textwidth]{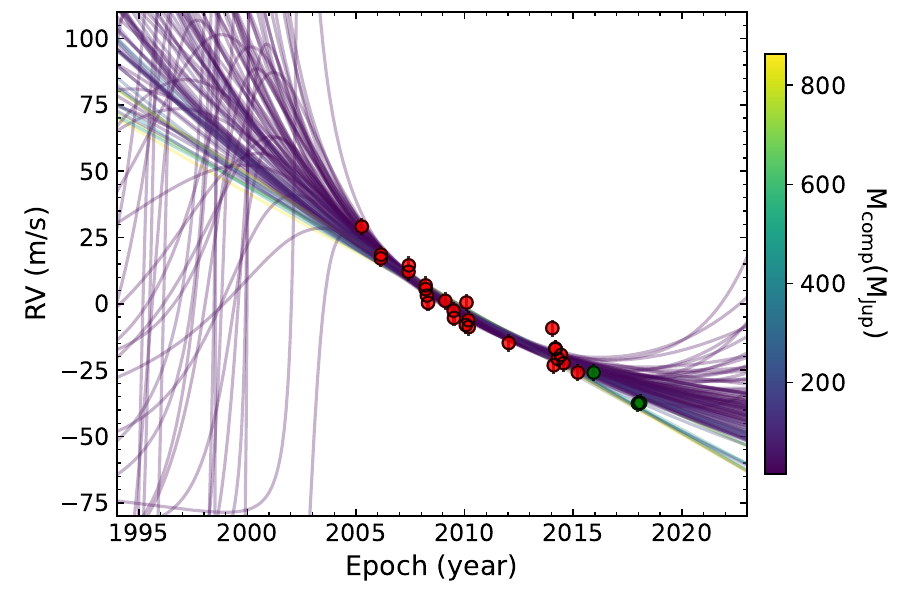}
    \includegraphics[width=0.5\textwidth]{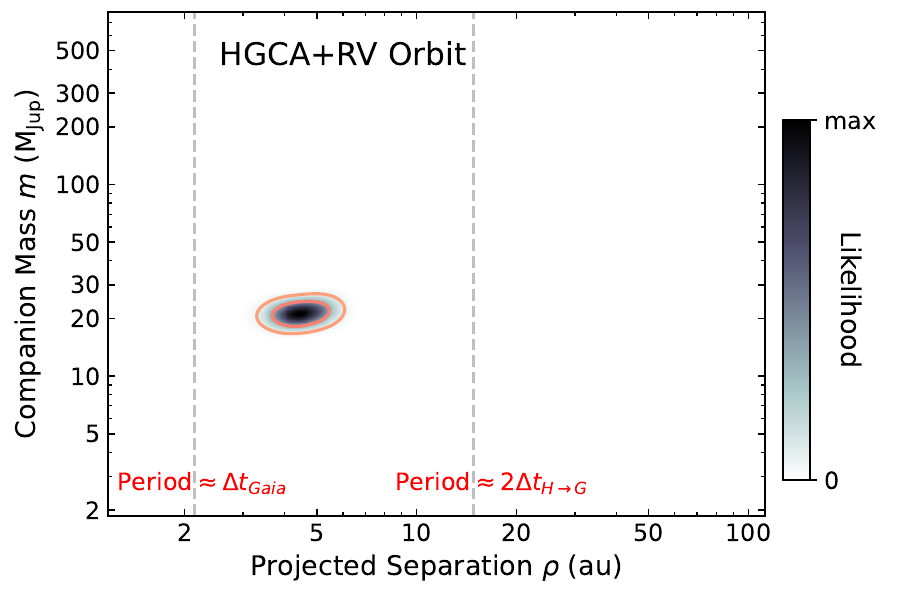}
    \includegraphics[width=0.5\textwidth]{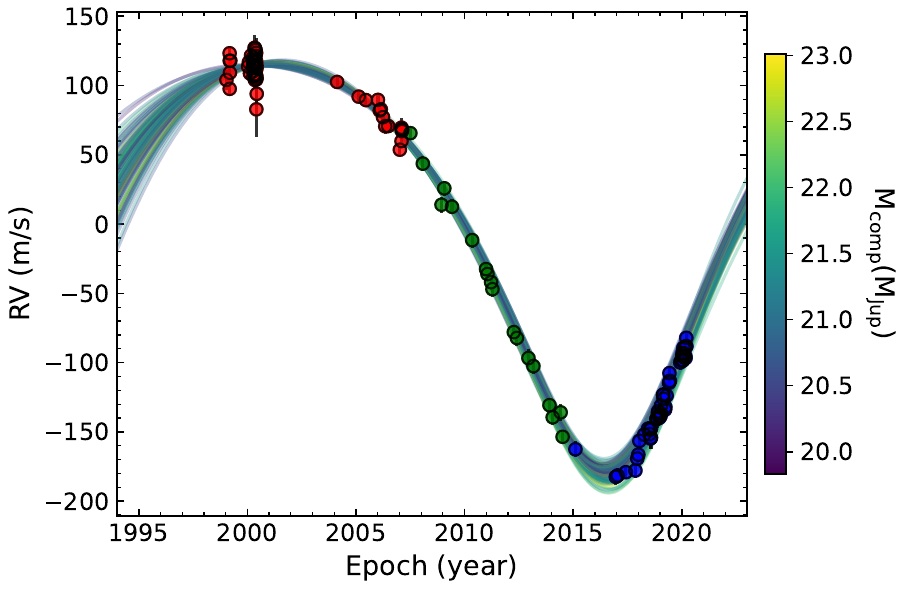}
    \caption{Left: projected separation consistent with an assumed 13$\sigma$ significant proper motion difference in the Hipparcos-Gaia Catalog of Accelerations \citep{Brandt_2021} assuming no additional data (top), a ten-year radial velocity trend (middle), and most of a radial velocity orbit (bottom).  Right: the radial velocity measurements adopted, color-coded by different assumed instruments, together with random orbits from the MCMC orbit fitting code {\tt orvara}.  The actual radial velocity points were those published for HIP 58289 \citep{Trifonov+Zechmeister+Kaminski+etal_2020} for the trend and HD 92987 \citep{Rickman+Segransan+Marmier+etal_2019} for the orbit.  }
    \label{fig:astrometric_prediction}
\end{figure*}

With only an astrometric acceleration and no additional data, we can only apply a joint constraint to the companion's mass and separation at a given epoch.  Figure \ref{fig:astrometric_prediction} shows an example of the constraints that are possible; similar figures appear in, e.g., \cite{Kervella+Arenou+Mignard+etal_2019} and \cite{Franson+Bowler+Zhou+etal_2023}.  The three left-hand panels of Figure \ref{fig:astrometric_prediction} include the cases where only absolute astrometry is available (top), where a ten-year RV trend is also available (middle), and where radial velocity measurements cover much of an orbit (bottom).  They show the probability density of a companion having a given mass and projected separation; the contours enclose 68\% and 95\% of the probability integrated over the region shown.  The right panels of Figure \ref{fig:astrometric_prediction} show the radial velocity data assumed and sample orbital fits from the MCMC orbit fitting code {\tt orvara} \citep{Brandt+Dupuy+Li+etal_2021}.  The figure adopts the astrometry measured for HIP 21152, which shows acceleration at $\approx$1$3\sigma$ significance due to a recently imaged brown dwarf companion \citep{Bonavita+Fontanive+Gratton+etal_2022,Kuzuhara+Currie+Takarada+etal_2022,Franson+Bowler+Bonavita+etal_2023}.  Figure \ref{fig:astrometric_prediction} shows the projected separations at 2022.0 that are consistent with the relevant assumed data.  

The top panel of Figure \ref{fig:astrometric_prediction} shows three distinct regimes.  At wide separations, where the period is more than twice the total astrometric baseline of $\approx$25 years, absolute astrometry primarily measures an acceleration and constrains the quantity $m/r^2$.  The tail to higher masses (above the highest density region on the plot) arises from geometries where $\sin \phi$ in Equation \eqref{eq:a_perp} is small, i.e., the companion is far away from its host along the line of sight.  The high likelihood in the top panel of Figure \ref{fig:astrometric_prediction} extends to very high companion masses and very large separations.  At a sufficiently wide separation, the companion would be massive enough to be comparably bright to its host.  In this case, it should be detected separately in Gaia, and a nondetection of the companion in Gaia can be used to set an upper limit on its mass and separation.  

The second regime in the top panel of Figure \ref{fig:astrometric_prediction} lies at intermediate separations.  These separations, via Kepler's Third Law, correspond to periods between the time baseline covered by Gaia alone and twice the baseline between Hipparcos and Gaia.  The six orbits shown in Figure \ref{fig:propermotion_demo} span this window.  
In this window the companion mass $m$ needed to account for the proper motion anomaly depends much more weakly on separation.  At these periods the star will cover a large fraction of an orbit, or even multiple orbits, between the two position measurements.  As a result, the position difference between the Hipparcos and Gaia measurements is much lower than the orbital speed times the time interval.   The proper motion anomaly, instead of measuring an astrometric acceleration, measures something much closer to an orbital speed, which scales as $m/\sqrt{a}$.  

The final regime is at close separations.  When the orbital period approaches the duration of the Gaia mission, a constant velocity fit is no longer a good model to the sky path, and the orbit must be fit with the individual Gaia measurements.  These fits are available for a number of stars in Gaia DR3 \citep{Halbwachs+Pourbaix+Arenou+etal_2023,Holl+Sozzetti+Sahlmann+etal_2023}, and will be available for many more stars in the future.  At these short orbital periods a five-parameter fit becomes a poor model for the sky path seen in Gaia.  The renormalized unit weight error (RUWE), a measure of the goodness-of-fit in Gaia, can provide an indication of these situations.

As noted in Section \ref{sec:longperiod}, if radial velocity data are also available, the parameter space shown in the top panel of Figure \ref{fig:astrometric_prediction} will be better constrained.  A measurement of a radial velocity trend fixes the angle between the separation and the line-of-sight, collapsing the cloud of points around $m \sim r^2$ to the line.  A radial velocity trend can also rule out shorter orbital periods.  The middle panel of Figure \ref{fig:astrometric_prediction} shows this case.  The actual radial velocity measurements are HARPS data taken from HIP 58289 \citep{Trifonov+Zechmeister+Kaminski+etal_2020}; the two colors shown in the right middle panel of Figure \ref{fig:astrometric_prediction} indicate two different instruments with different zero points.  In this case they are before and after a fiber replacement in HARPS.  

The lower panel of Figure \ref{fig:astrometric_prediction} shows the constraints that are possible if radial velocity monitoring covers most of an orbit.  The actual radial velocities in this case are CORALIE data taken from HD 92987 \citep{Rickman+Segransan+Marmier+etal_2019}, with three different time periods during which the instrument was stable.  The radial velocity orbit constrains the period directly.  The curves on the lower-right panel show random orbits consistent with the data as determined using the MCMC orbit fitting code {\tt orvara}.

The orbital constraints shown in the lower panel of Figure \ref{fig:astrometric_prediction} assume that a good constraint on the mass of the host star is available.  Such a constraint enables the semimajor axis to be inferred from the orbital period via Kepler's Third Law.  Absolute astrometry then constrains the orientation of the orbit and determines the companion mass and projected separation.  This collapses the cloud of points in the left panels to nearly a point.  This is similar to the situation with the discovery of Neptune, where its mass and position could be accurately predicted in advance of imaging.

\section{Future Prospects} \label{sec:futureprospects}

Gaia continues to collect data and improve its precision.  Gaia DR2 represented about two years of observations.  The precision of the Gaia DR2 proper motions was lower than the precision of the long-term Hipparcos-Gaia proper motions, typically by a factor of a few \citep{Brandt_2018}.  With the release of Gaia DR3, representing nearly three years of data together with improved data processing \citep{Lindegren+Klioner+Hernandez+etal_2020}, the precision of Gaia proper motions is now comparable to the precision of the long-term proper motions \citep{Brandt_2021,Kervella+Arenou+Thevenin_2022}.  However, this does not mean that Hipparcos no longer adds value.  In particular, the precision of Gaia alone in measuring astrometric accelerations of bright stars remains far inferior to that enabled by including Hipparcos positions.  

To understand the continuing importance of Hipparcos, it is necessary to understand how Gaia's precision scales with its observing baseline.  The benefit of additional observing time manifests very differently for measurements of positions, proper motions, and astrometric accelerations. 
The precision of measuring a position scales with the number of measurements $n$ as $\sqrt{n}$.  Assuming these measurements to be evenly spaced in time, the astrometric precision of a survey in position scales as $\sqrt{t}$.  Because parallactic motion is periodic with a period of one year, its precision scales similarly for a multi-year survey that spans many parallactic periods.  In both cases---position and parallax---a mission that lasts twice as long should have uncertainties that are lower by a factor $\approx$$\sqrt{2}$.  

\begin{figure}
    \includegraphics[width=\linewidth]{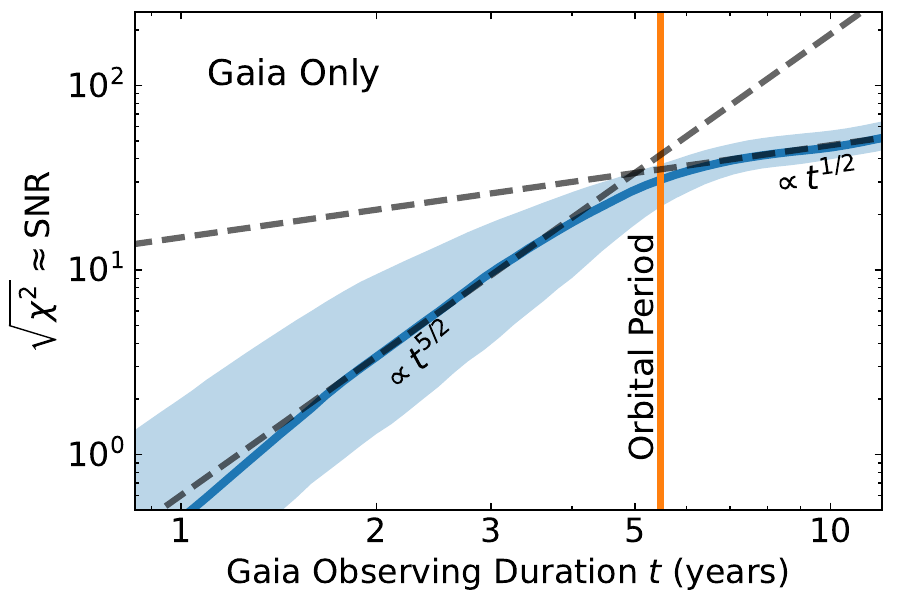}
    \includegraphics[width=\linewidth]{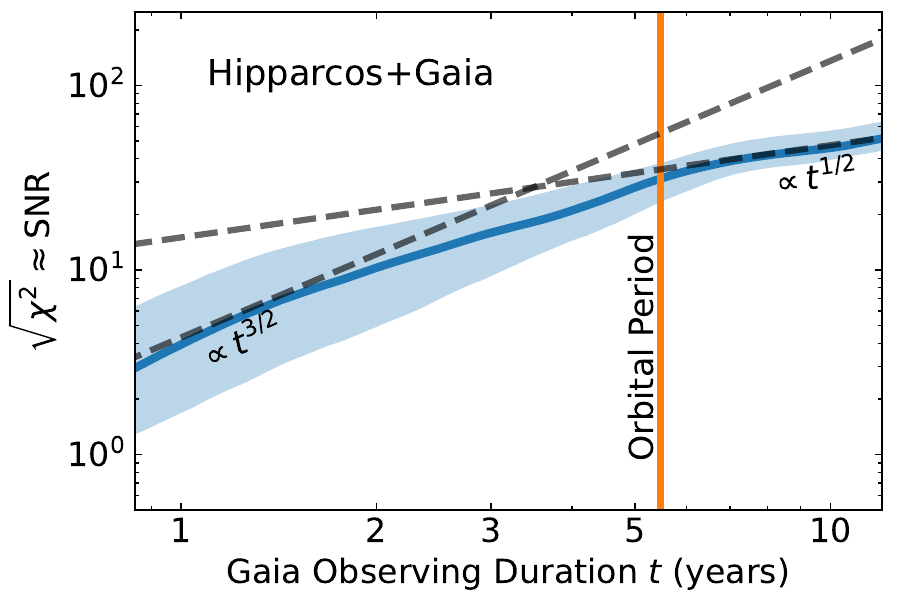}
    \includegraphics[width=\linewidth]{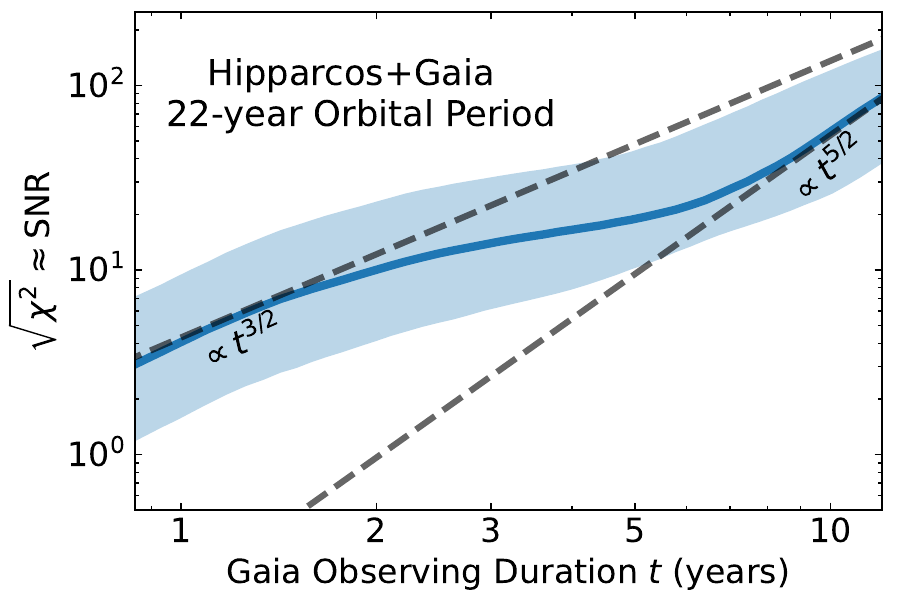}
    \caption{Illustration of the different companion sensitivity scalings with Gaia baseline that are possible depending on orbital period and the existence of Hipparcos data.  The shaded region encloses 90\% of random orbits with eccentricity 0.5.  Top: with Gaia only, acceleration sensitivity scales as $t^{5/2}$ until the baseline reaches the orbital period, then as $t^{1/2}$.  Middle: the initial scaling is $t^{3/2}$ because the relevant measurement from Gaia is a proper motion.  Bottom: for a long period and a sufficiently long-duration Gaia mission, the Hipparcos constraint finally ceases to add much and the sensitivity scales as $t^{5/2}$.}
    \label{fig:scaling_futuregaia}
\end{figure}

The precision of a velocity or acceleration measurement scales very differently with observing baseline.  A star that moves uniformly across the sky will have a position difference that scales as the survey duration.  The longer a survey continues, the further the star has moved, and the easier this motion is to see.  While the precision of a position measurement scales as with the mission baseline $t$ as $\sqrt{t}$, the displacement of a star across the mission adds another factor of $t$ for a total scaling of $t^{3/2}$.  

For astrometric accelerations the scaling with mission baseline is even more extreme.  The position displacement of an accelerating source after time $t$ scales as $t^2$.  This adds an additional factor of $t$ beyond the scaling of proper motion precision with mission baseline, for a total scaling of $t^{5/2}$.  

Figure \ref{fig:scaling_futuregaia} shows how these general scaling relations impact the sensitivity of Gaia now and in the future to two sample companions: one at an orbital period of 2000 days ($\approx$5.5 years) for the top two panels, and one with a 22-year orbital period for the bottom panel.  Both companions are assumed to have an orbital eccentricity of 0.5.  The shaded regions include 90\% of orbits at random phases and orientations.  

The top panel of Figure \ref{fig:scaling_futuregaia} assumes that only Gaia data is used; the middle and bottom panels assume that Hipparcos also provides a position measurement 25 years in the past.  The vertical axis in all cases is the significance of deviation from a five-parameter sky path as measured by the square root of the $\chi^2$ improvement from using the correct orbital model.  It approximates the signal-to-noise ratio of a detection of non-uniform space motion.  The vertical scale is arbitrary: it depends on the astrometric precision, the distance of the star from Earth, and the mass of the companion.   

The top panel of Figure \ref{fig:scaling_futuregaia} uses only Gaia data.  In this case, the sensitivity initially scales nearly as $t^{5/2}$ as Gaia increases its observing baseline.  This may be intuitively understood as being due to the scaling of acceleration sensitivity with mission baseline.  As the observing baseline exceeds the orbital period, the scaling approaches the $t^{1/2}$ sensitivity seen for position and parallax.

The middle panel of Figure \ref{fig:scaling_futuregaia} is similar to the top panel, but includes a Hipparcos position measurement.  At small Gaia observing durations, Gaia itself is insensitive to accelerations.  An acceleration must be inferred from a difference between the Gaia proper motion and the long-term proper motion given by the Hipparcos-Gaia positional shift.  Since it is Gaia's proper motion measurement that determines the sensitivity, its precision scales as $t^{3/2}$.  As before, as the total Gaia baseline exceeds the orbital period, the precision of the joint astrometric data sets scales as $\sqrt{t}$.

The bottom panel of Figure \ref{fig:scaling_futuregaia} adopts a longer-period companion.  At short orbital periods, Gaia is again insensitive to accelerations on its own, so the sensitivity of the combined astrometric data set scales as $t^{3/2}$.  There is then a plateau where the scaling is closer to $\sqrt{t}$.  The precision of the Gaia proper motion at this point has exceeded the precision of the Hipparcos-Gaia long-term proper motion, while the sensitivity of Gaia itself to accelerations has not yet reached the precision of the difference between the two proper motions.  Then, at slightly longer baselines, Gaia itself is more sensitive to accelerations than the proper motion difference with Hipparcos.  At these mission durations Hipparcos finally ceases to be relevant and Gaia's sensitivity scales as $t^{5/2}$.

Hipparcos currently provides position measurements for a little more than 100,000 stars brighter than $\approx$11 magnitude.  For the millions of stars fainter than this, pre-Gaia position measurements were much lower precision than Hipparcos, e.g.~from digitized photographic plates \citep{Lehtinen+Prusti+deBruijne+etal_2023} or from the Hipparcos star mapper \citep[the Tycho-2 catalog,][]{Hog+Fabricius+Makarov+etal_2000}.  Without a sufficiently precise astrometric data point in the past, the scaling with Gaia baseline shown in the top panel of Figure \ref{fig:scaling_futuregaia} will apply.  As Gaia completes its extended mission and ultimately releases $\approx$10 years of data it will conduct an extensive census of planet masses and orbits with periods out to $\approx$10 years.  Assuming Gaia data to be statistically well-behaved, this future Gaia data set will be much more sensitive than Gaia DR3.

The middle and lower panels of Figure \ref{fig:scaling_futuregaia} show that Hipparcos will continue to have value well into the future.  The same principle also applies to future astrometric measurements: a position measurement years into the future can add a large amount of value to Gaia astrometry.  This measurement could be provided by a future mission intended for precise astrometry, like a proposed near-infrared successor to Gaia \citep{Hobbs+Hog_2018,McArthur+Hobbs+Hog+etal_2019}\footnote{\url{https://www.astro.lu.se/GaiaNIR}}.  Such a mission would also enable a cross-calibration of Gaia and significant improvements in systematics.  A future position measurement could even be provided by another mission like HST, JWST, or the Roman space telescope assuming that it can provide high-precision astrometric measurements for relatively bright stars \citep{Melchior+Spergel+Lanz_2018,WFIRST_Astrometry_2019}.  

\section{Conclusions} \label{sec:conclusions}

This tutorial has reviewed the application of absolute astrometry, particularly the combination of Hipparcos and Gaia, to find and weigh faint companions of nearby stars.  I first reviewed the basic properties of absolute astrometry and the sky paths followed by stars traveling on inertial trajectories.  I then summarized the effects of a companion on this astrometric sky path and the currently available astrometric data sets that can detect the influence of an unseen companion.  In order to be useful to fit orbits, these data sets must have well-understood and well-calibrated statistical properties.  Orbit fitting tends to begin by writing down a likelihood, which implicitly assumes that all data have Gaussian errors of known variance, and often assumes that the data have zero covariance.  I then reviewed the intuition for how dynamical masses can be precisely measured and how the possible constraints depend the properties of a binary system. 

Astrometry from Hipparcos and Gaia has fundamentally changed the kinds of systems that are amenable to dynamical mass measurements.  Dynamical masses from tracing out orbits on the sky or from measuring radial velocities typically require data to cover most or all of an orbit.  For long-period systems, this has meant that masses are only possible for those systems with data going back decades or even centuries.  The combination of the Hipparcos and Gaia data sets have enabled mass measurements of these long-period systems by measuring their accelerations in an inertial reference frame.  Absolute astrometry of systems on century-long orbits is now providing some of the most precisely measured masses of brown dwarfs, white dwarfs, and low-mass stars.  

As Gaia continues to observe, our sensitivity to faint companions and our ability to measure masses will steadily improve.  Gaia's extended mission will ultimately include $\sim$10 years of data, which will place its highest astrometric sensitivity to orbits near this period.  For longer-period systems, Gaia's sensitivity to acceleration will improve rapidly and will enable dynamical mass measurements for an ever-increasing number of nearby systems.  The same principles that have allowed Hipparcos to add so much value to Gaia---the addition of a position measurement far displaced in time---mean that a similar strategy in the future could continue to extend Gaia's sensivity to ever wider and lower-mass companions.

\bibliographystyle{aasjournal}
\bibliography{refs}

\end{document}